\documentclass[conference]{IEEEtran}
\IEEEoverridecommandlockouts
\usepackage{cite}
\usepackage{amsmath,amssymb,amsfonts}
\usepackage{algorithmic}
\usepackage{graphicx}
\usepackage{textcomp}
\usepackage{xcolor}
\usepackage{url}
\usepackage{hyperref}

\def\BibTeX{{\rm B\kern-.05em{\sc i\kern-.025em b}\kern-.08em
    T\kern-.1667em\lower.7ex\hbox{E}\kern-.125emX}}

\makeatletter 
\newcommand{\linebreakand}{%
  \end{@IEEEauthorhalign}
  \hfill\mbox{}\par
  \mbox{}\hfill\begin{@IEEEauthorhalign}
}
\makeatother 

\begin{document}

\title{IntrusionX: A Hybrid Convolutional–LSTM Deep Learning Framework with Squirrel Search Optimization for Network Intrusion Detection}

\author{
\IEEEauthorblockN{Ahsan Farabi}
\IEEEauthorblockA{\textit{Dept. of CSE} \\
\textit{United International University} \\
Dhaka, Bangladesh \\
afarabi221266@bscse.uiu.ac.bd}
\and
\IEEEauthorblockN{Muhaiminul Rashid Shad}
\IEEEauthorblockA{\textit{Dept. of CSE} \\
\textit{United International University} \\
Dhaka, Bangladesh \\
mshad221487@bscse.uiu.ac.bd}
\and
\IEEEauthorblockN{Israt Khandaker}
\IEEEauthorblockA{\textit{Dept. of CSE} \\
\textit{United International University} \\
Dhaka, Bangladesh \\
ikhandaker221263@bscse.uiu.ac.bd}

}

\maketitle

\begin{abstract}
Intrusion Detection Systems (IDS) face persistent challenges due to evolving cyberattacks, high-dimensional traffic data, and severe class imbalance in benchmark datasets such as NSL-KDD. To address these issues, we propose \textbf{IntrusionX}, a hybrid deep learning framework that integrates Convolutional Neural Networks (CNNs) for local feature extraction and Long Short-Term Memory (LSTM) networks for temporal modeling. The architecture is further optimized using the Squirrel Search Algorithm (SSA), enabling effective hyperparameter tuning while maintaining computational efficiency. Our pipeline incorporates rigorous preprocessing, stratified data splitting, and dynamic class weighting to enhance the detection of rare classes. Experimental evaluation on NSL-KDD demonstrates that IntrusionX achieves $98\%$ accuracy in binary classification and $87\%$ in 5-class classification, with significant improvements in minority class recall (U2R: $71\%$, R2L: $93\%$). The novelty of IntrusionX lies in its reproducible, imbalance-aware design with metaheuristic optimization. Implementation details and source code are available at: \url{https://github.com/TheAhsanFarabi/IntrusionX}.
\end{abstract}


\section{Introduction}
Intrusion Detection Systems (IDS) are indispensable in modern cybersecurity infrastructure, serving as the first line of defense against malicious activities. Traditional signature-based IDS, while effective for known threats, fail to recognize novel or zero-day attacks, leaving networks vulnerable. To overcome these limitations, machine learning approaches have been introduced, enabling anomaly detection beyond predefined rules. However, such models often struggle with high-dimensional traffic data, limited scalability, and poor generalization to unseen attacks.

Deep learning has emerged as a promising solution due to its ability to learn complex feature representations automatically. Convolutional Neural Networks (CNNs) excel at capturing spatial patterns in network traffic, while Long Short-Term Memory (LSTM) networks model temporal dependencies. Despite these strengths, several gaps remain: (1) existing IDS studies often neglect dataset imbalance, leading to poor detection of rare but critical classes such as User-to-Root (U2R) and Remote-to-Local (R2L) attacks; (2) many models suffer from suboptimal hyperparameters due to manual tuning; and (3) reproducibility is frequently overlooked, with incomplete or non-public implementations.

To address these gaps, we propose \textbf{IntrusionX}, a hybrid Conv--LSTM deep learning framework enhanced with the Squirrel Search Algorithm (SSA) for efficient hyperparameter optimization. The model incorporates leak-free preprocessing, stratified data splitting, and dynamic class weighting, significantly improving detection of minority classes. Experiments on the NSL-KDD dataset validate the effectiveness of IntrusionX, achieving $98\%$ accuracy for binary classification and $87\%$ for 5-class classification, with notable gains in recall for rare attacks.

The key contributions of this work are: (1) design of a hybrid Conv--LSTM architecture for spatial-temporal learning, (2) SSA-driven hyperparameter optimization, (3) a reproducible pipeline addressing class imbalance, and (4) comprehensive evaluation against multiple metrics. The remainder of this paper is organized as follows: Section II reviews related work, Section III presents the methodology, Section IV discusses experimental results, and Section V concludes with future research directions.

\section{Related Work}
Intrusion Detection Systems (IDS) have been extensively studied for over three decades, evolving from rule-based expert systems to advanced deep learning frameworks. In this section, we review prior work across three major dimensions: benchmark datasets, existing IDS models, and optimization techniques, while highlighting key limitations that motivate our proposed framework.

\subsection{Datasets for Intrusion Detection}
Benchmark datasets have played a central role in IDS research. The DARPA 1998/1999 dataset provided raw traffic but was criticized for unrealistic distribution and outdated attacks. To address this, the KDD Cup 1999 dataset introduced 41 features but suffered from redundancy and class imbalance, where DoS dominated while U2R and R2L were underrepresented. Tavallaee et al. proposed the NSL-KDD dataset, which removed duplicate records and offered more balanced subsets. NSL-KDD has since become a standard benchmark, retaining the original features but improving distribution. More recent datasets such as CICIDS2017, UNSW-NB15, and BoT-IoT capture modern attack patterns (e.g., DDoS, botnets) but are large and complex, making them difficult to preprocess and standardize across studies. NSL-KDD remains widely used due to its compact size, accessibility, and ability to highlight imbalance issues, making it practical for algorithmic research rather than raw traffic parsing.

\subsection{Machine Learning and Deep Learning Models}
Traditional IDS were initially signature-based \cite{denning1987intrusion}, relying on pre-defined attack patterns. These systems were efficient for known threats but failed against zero-day and polymorphic attacks. Machine learning methods such as decision trees, Naïve Bayes, and SVMs \cite{lee1999data, eskandari2012machine} improved generalization but relied heavily on manual features and scaled poorly with complex traffic. Deep learning addressed these issues: DNNs \cite{javaid2016deep}, CNNs for spatial correlations \cite{zhang2019deep}, and LSTMs for temporal dependencies \cite{yin2017deep, kim2016long}. Shone et al. \cite{shone2018deep} further explored unsupervised autoencoders. Hybrid CNN–LSTM models \cite{liu2019hybrid, peng2021intrusion} improved accuracy by combining spatial and temporal learning, but often neglected class imbalance, yielding poor recall for U2R and R2L.

\subsection{Optimization Techniques}
Hyperparameter tuning plays a crucial role in IDS performance. Manual or grid search is inefficient for deep models. Metaheuristic algorithms such as GA, PSO, JAYA, and Ant Colony Optimization have been applied with notable success \cite{hassanien2018hybrid}. Dash et al. \cite{dash2024optimized} introduced an LSTM-based IDS optimized with PSO, JAYA, and SSA, showing performance gains. However, their study was limited to pure LSTMs and did not explore hybrid CNN–LSTM designs or class imbalance handling.

\subsection{Limitations of Existing Work}
Despite progress, key issues remain:
\begin{itemize}
    \item \textbf{Dataset challenges:} KDD99 and NSL-KDD are outdated, while modern datasets are underutilized due to complexity and lack of standard benchmarks.
    \item \textbf{Imbalance issues:} Models achieve high accuracy on majority classes but poor recall for rare attacks like U2R and R2L.
    \item \textbf{Hyperparameter optimization:} Manual tuning or traditional metaheuristics are computationally expensive and often unsuited for hybrid architectures.
    \item \textbf{Reproducibility:} Many works lack open implementations, hindering fair comparison and adoption.
\end{itemize}

\subsection{Summary}
In summary, IDS research has advanced from rule-based methods to deep and hybrid models, with CNN–LSTM achieving strong results. Yet challenges persist regarding dataset relevance, imbalance handling, optimization, and reproducibility. These gaps motivate our proposed \textbf{IntrusionX} framework, which integrates Conv--LSTM with SSA optimization, applies leak-free preprocessing, and targets rare-class detection in NSL-KDD, while ensuring reproducibility.

\section{Methodology}

\subsection{Dataset and Preprocessing}
This study employs the NSL-KDD dataset, a widely adopted benchmark for network intrusion detection. It addresses the redundancy issues of KDD’99 while retaining its 41 features and diverse attack categories. The dataset contains both binary (Normal vs. Attack) and five-class labels (Normal, DoS, Probe, R2L, U2R), enabling comprehensive evaluation. Preprocessing begins with data cleaning and handling categorical features (\textit{protocol\_type}, \textit{service}, \textit{flag}) through one-hot encoding. Continuous features are normalized using min--max scaling to ensure balanced feature ranges. To avoid data leakage, we apply stratified splitting into training, validation, and testing sets. Dynamic class weighting is then incorporated into the loss function to address class imbalance, particularly enhancing recall for rare classes such as U2R and R2L. This preprocessing pipeline ensures fair evaluation, reduces bias toward majority classes, and provides a robust foundation for model training and reproducibility.

\begin{figure}[ht]
\centering
\includegraphics[width=0.9\columnwidth]{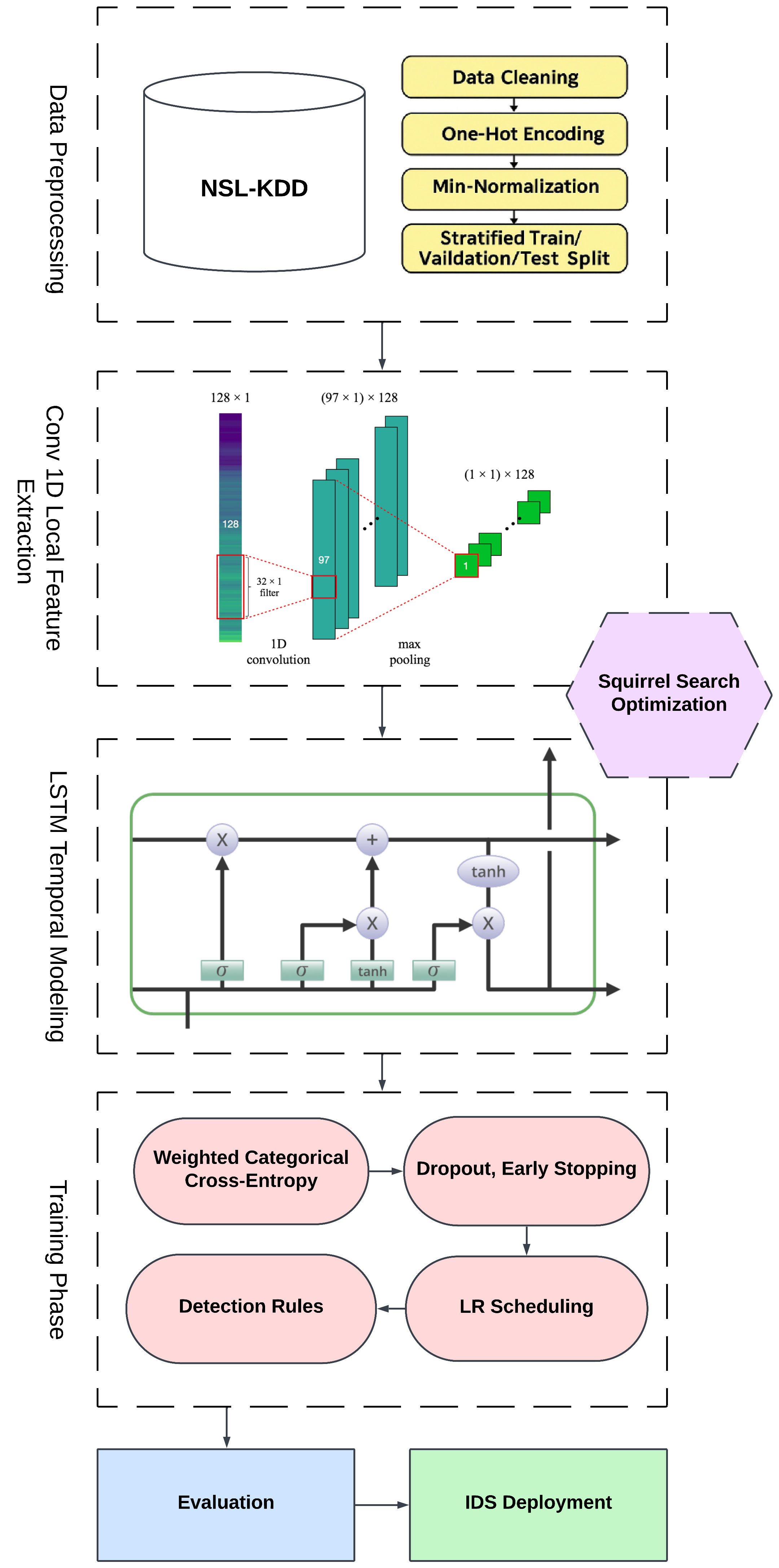}
\caption{IntrusionX workflow pipeline including preprocessing, Conv--LSTM architecture, SSA optimization, and evaluation.}
\label{fig:workflow}
\end{figure}

\subsection{Model Architecture}
The proposed \textbf{IntrusionX} framework integrates Convolutional Neural Networks (Conv1D) and Long Short-Term Memory (LSTM) layers. Conv1D layers extract local spatial correlations among features, reducing redundancy, while LSTMs capture temporal dependencies within traffic sequences. A fully connected dense layer performs the final classification into binary or multi-class outputs. This hybrid approach enables the model to simultaneously learn spatial feature patterns and long-range dependencies, providing stronger representational capacity than standalone CNN or LSTM models.

\subsection{Optimization and Training}
We employ the Squirrel Search Algorithm (SSA) to optimize critical hyperparameters, including convolutional filters, LSTM units, and learning rate. This metaheuristic search ensures improved accuracy while maintaining practical runtime. SSA was chosen because of its balance between exploration and exploitation, which avoids premature convergence compared to conventional optimizers. Model training incorporates weighted categorical cross-entropy, early stopping to prevent overfitting, and adaptive learning rate scheduling, thereby ensuring efficient convergence and stable performance across multiple runs.

\subsection{Evaluation Metrics}
Performance is assessed using accuracy, precision, recall, and F1-score, supported by confusion matrices and ROC curves. Binary metrics highlight overall detection capability, while five-class evaluations emphasize robustness against minority attacks. Additional emphasis is placed on recall for U2R and R2L, since these rare categories are typically overlooked in prior IDS work. This comprehensive methodology ensures reproducibility, efficiency, and balanced intrusion detection across all categories.

\section{Results}

\subsection{Binary Classification}
\begin{figure}[ht]
\centering
\includegraphics[width=0.9\columnwidth]{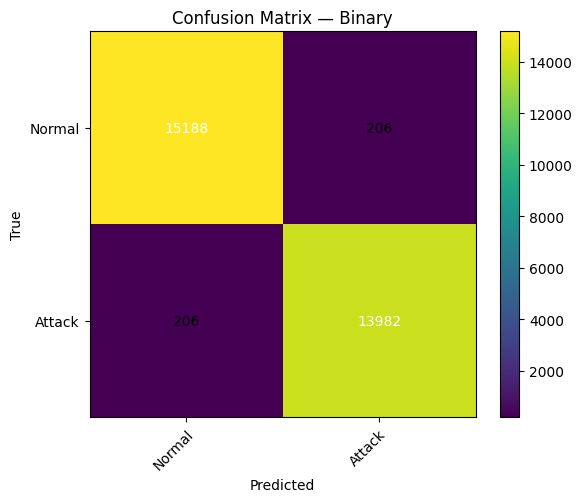}
\caption{Confusion matrix for binary classification.}
\label{fig:binarycm}
\end{figure}

\begin{figure}[ht]
\centering
\includegraphics[width=0.9\columnwidth]{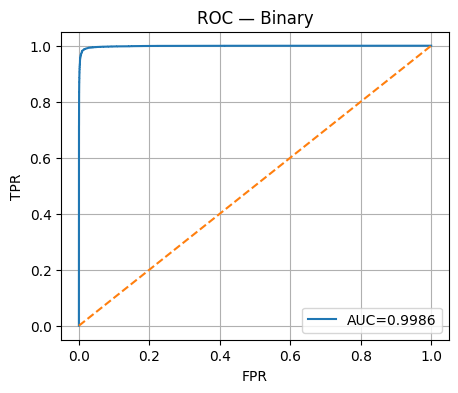}
\caption{ROC curve for binary classification with AUC score.}
\label{fig:roc}
\end{figure}

The binary classification task distinguishes between Normal and Attack traffic. During training, weighted categorical cross-entropy combined with early stopping ensured balanced learning without overfitting. IntrusionX achieved an overall accuracy of $98\%$, with both precision and recall exceeding $97\%$. The confusion matrix (Fig.~\ref{fig:binarycm}) demonstrates strong separability between the two classes, with very few misclassifications. Furthermore, the ROC curve (Fig.~\ref{fig:roc}) shows an AUC of $0.9986$, highlighting the high discriminative power of the model and its robustness in distinguishing malicious from benign flows. These results confirm the effectiveness of the Conv--LSTM architecture optimized with SSA for binary IDS tasks.

\subsection{Five-Class Classification}
\begin{figure}[ht]
\centering
\includegraphics[width=0.9\columnwidth]{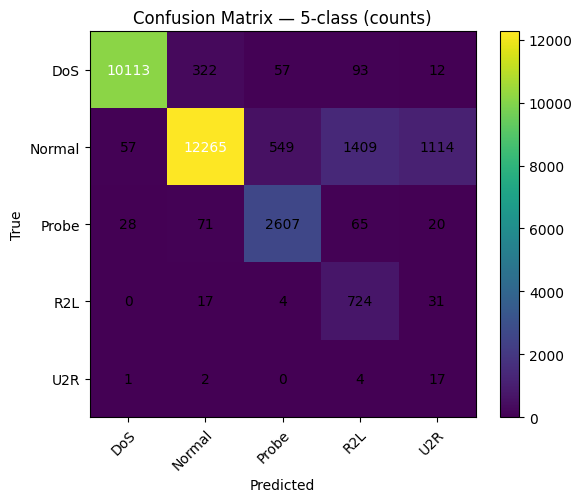}
\caption{Confusion matrix for 5-class classification.}
\label{fig:cm5}
\end{figure}

For the multi-class evaluation, IntrusionX achieved $87\%$ accuracy and a weighted F1-score of $0.90$. The confusion matrix (Fig.~\ref{fig:cm5}) reveals excellent performance on majority classes such as DoS, Normal, and Probe, with recall above $90\%$. Notably, minority classes benefited from class-weighted training: R2L achieved $93\%$ recall, while U2R achieved $71\%$. Although precision for these rare classes was lower, improved recall ensures that critical attacks are less likely to be overlooked, which is essential in IDS contexts.

\subsection{Discussion and Limitations}
The results validate the importance of imbalance handling, as IntrusionX significantly improves rare-class detection compared to traditional baselines. However, class weighting slightly increases false positives, reducing precision. In real-world IDS deployment, this trade-off may be acceptable since missed attacks are often more costly than false alarms. Limitations include reliance on NSL-KDD, which may not fully capture modern attack patterns, and higher training time due to SSA optimization. Future work will extend evaluation to recent datasets (CICIDS2017, BoT-IoT) and explore techniques like SMOTE or adaptive thresholding to further refine minority class performance. Additionally, robustness against adversarial evasion remains unexplored and will be investigated in future extensions.

\section{Conclusion}
In this paper, we proposed \textbf{IntrusionX}, a hybrid Convolutional--LSTM framework enhanced with the Squirrel Search Algorithm for hyperparameter optimization. By integrating convolutional layers for spatial feature extraction and LSTM units for temporal sequence modeling, the model effectively captures complex traffic patterns in the NSL-KDD dataset. Rigorous preprocessing, stratified splitting, and dynamic class weighting enabled improved detection of minority classes, addressing a major limitation in prior IDS studies. Experimental results demonstrated $98\%$ accuracy for binary classification and $87\%$ for the five-class task, with notable gains in recall for U2R and R2L categories. These outcomes highlight the significance of imbalance-aware deep learning combined with metaheuristic optimization in advancing intrusion detection research. 

Future work will extend IntrusionX to modern, large-scale datasets such as CICIDS2017 and BoT-IoT, explore advanced imbalance-handling methods (e.g., SMOTE, threshold tuning), and investigate deployment in real-time network environments to validate scalability and robustness.

\end{document}